\documentclass[%
superscriptaddress,
showpacs,
 amsmath,amssymb,
 aps,
prl,
twocolumn,
]{revtex4-1}

\usepackage{array}
\usepackage{wrapfig}

\usepackage{graphicx}
\usepackage{dcolumn}
\usepackage{bm}
\usepackage{amsmath}
\usepackage{amssymb}
\usepackage{graphics}
\usepackage{epsfig}
\usepackage{natbib}
\usepackage{verbatim} 

\usepackage{color}

\usepackage{siunitx} 

\newcommand{\muSR}{$\mu$SR }
\newcommand{\FeSeS}{FeSe$_\mathrm{1\text{-}x}$S$_\mathrm{x}$ }
\newcommand{\FeSeSd}{Fe$_{1+\delta}$Se$_\mathrm{1\text{-}x}$S$_\mathrm{x}$ }
\newcommand{\Tc}{$T_\mathrm{c}$ }
\newcommand{\TN}{$T_\mathrm{N}$ }
\newcommand{\muSRp}{$\mu$SR}

\newcommand{\Tcp}{$T_\mathrm{c}$}
\newcommand{\TNp}{$T_\mathrm{N}$}

\begin{document}


\title{Extended Magnetic Dome Induced by Low Pressures in Superconducting FeSe$_\mathrm{1\text{-}x}$S$_\mathrm{x}$}

\author{S.~Holenstein}
\email{stefan.holenstein@psi.ch}
\affiliation{Laboratory for Muon Spin Spectroscopy, Paul Scherrer
Institute, CH-5232 Villigen PSI, Switzerland}
\affiliation{Physik-Institut der Universit\"at Z\"urich,
Winterthurerstrasse 190, CH-8057 Z\"urich, Switzerland}
\author{J.~Stahl}
\affiliation{Department Chemie, Ludwig-Maximilians-Universit\"at M\"unchen,
Butenandtstr. 5-13 (D), 81377 M\"unchen, Germany}
\author{Z.~Shermadini}
\affiliation{Laboratory for Muon Spin Spectroscopy, Paul Scherrer
Institute, CH-5232 Villigen PSI, Switzerland}
\author{G.~Simutis}
\affiliation{Laboratory for Muon Spin Spectroscopy, Paul Scherrer
Institute, CH-5232 Villigen PSI, Switzerland}
\author{V.~Grinenko}
\affiliation{Institute of Solid State and Materials Physics, TU Dresden, DE-01069
Dresden, Germany}
\affiliation{Institute for Metallic Materials, Leibniz IFW Dresden, DE-01069 Dresden, Germany}
\author{D.~A.~Chareev}
\affiliation{RAS, Institute of Experimental Mineralogy, Chernogolovka 123456, Russia}
\affiliation{Ural Federal University, Ekaterinburg 620002, Russia}
\affiliation{Kazan Federal University, Kazan 420008, Russia}
\author{R.~Khasanov}
\affiliation{Laboratory for Muon Spin Spectroscopy, Paul Scherrer
Institute, CH-5232 Villigen PSI, Switzerland}
\author{J.-C.~Orain}
\affiliation{Laboratory for Muon Spin Spectroscopy, Paul Scherrer
Institute, CH-5232 Villigen PSI, Switzerland}
\author{A.~Amato}
\affiliation{Laboratory for Muon Spin Spectroscopy, Paul Scherrer
Institute, CH-5232 Villigen PSI, Switzerland}
\author{H.-H.~Klauss}
\affiliation{Institute of Solid State and Materials Physics, TU Dresden, DE-01069
Dresden, Germany}
\author{E.~Morenzoni}
\affiliation{Laboratory for Muon Spin Spectroscopy, Paul Scherrer
Institute, CH-5232 Villigen PSI, Switzerland}
\affiliation{Physik-Institut der Universit\"at Z\"urich,
Winterthurerstrasse 190, CH-8057 Z\"urich, Switzerland}
\author{D.~Johrendt}
\affiliation{Department Chemie, Ludwig-Maximilians-Universit\"at M\"unchen,
Butenandtstr. 5-13 (D), 81377 M\"unchen, Germany}
\author{H.~Luetkens}
\email{hubertus.luetkens@psi.ch}
\affiliation{Laboratory for Muon Spin Spectroscopy, Paul Scherrer
Institute, CH-5232 Villigen PSI, Switzerland}


\begin{abstract}
We report muon spin rotation ($\mu$SR) and magnetization measurements under pressure on Fe$_{1+\delta}$Se$_\mathrm{1\text{-}x}$S$_\mathrm{x}$ with x~$\approx 0.11$.
Above $p\approx0.6$~GPa we find microscopic coexistence of superconductivity with an extended dome of long range magnetic order that spans a pressure range between previously reported separated magnetic phases.
The magnetism initially competes on an atomic scale with the coexisting superconductivity leading to a local maximum and minimum of the superconducting $T_\mathrm{c}(p)$.
The maximum of $T_\mathrm{c}$ corresponds to the onset of magnetism while the minimum coincides with the pressure of strongest competition.
A shift of the maximum of $T_\mathrm{c}(p)$ for a series of single crystals with x up to 0.14 roughly extrapolates to a putative magnetic and superconducting state at ambient pressure for x~$\geq0.2$.

\end{abstract}

\maketitle


Unconventional superconductivity is usually obtained by suppressing a long range static magnetic order present in the parent compound. The suppression can be achieved via chemical doping or by application of pressure \cite{Mazin2010}. This is e.g. the case for cuprate \cite{Sanna2004} and iron based superconductors \cite{Uemura2009,Chen2009} as well as for heavy fermion \cite{Knebel2006,Uemura2009} and organic superconductors \cite{Sasaki2008}.
In iron pnictide compounds, the magnetic order is usually accompanied by a structural transition from a high temperature tetragonal to a low temperature orthorhombic phase \cite{Bohmer2017}. Surprisingly, the structurally simplest iron based superconductor FeSe is non-magnetic at ambient pressure but exhibits a structural phase transition which is associated with nematic order \cite{Hsu2008,Margadonna2008,McQueen2009a,Medvedev2009d,Margadonna2009c}. Under hydrostatic pressure, the nematic order is however suppressed, magnetic order emerges above \SI{0.9}{\giga\pascal} and the superconducting transition temperature \Tc increases from $\sim$\SI{8}{\kelvin} to $\sim$\SI{37}{\kelvin} at $\sim$\SI{7}{\giga\pascal} \cite{Bendele2010,Bendele2012,Medvedev2009d,Margadonna2009c,Mizuguchi2008,Garbarino2009,Masaki2009,Okabe2010,Miyoshi2014}. Despite the structural simplicity, the electronic properties of FeSe are highly non-trivial with reported Lifshitz transitions of the Fermi surface as a function of temperature \cite{Grinenko2018}, pressure \cite{Terashima2016c,Reiss2019}, and S substitution \cite{Coldea2019}. This rich phase diagram led to still ongoing discussions about the interplay of magnetism and nematicity and their respective influence on superconductivity \cite{Bohmer2017}.

In recent years, S substitution has come into focus as an additional tuning parameter for FeSe. The nematically ordered phase of \FeSeS is suppressed with increasing sulfur content and is no longer present for $\text{x}>0.17$ \cite{Tomita2015,Watson2015a,Ovchenkov2016,Abdel-Hafiez2016} where a nematic quantum critical point \cite{Hosoi2016}, a topological Lifshitz transition \cite{Coldea2019}, a reduction in electronic correlations \cite{Reiss2017}, and a change in the superconducting pairing state \cite{ Hanaguri2018} and gap structure \cite{Sato2018} are observed. Under high hydrostatic pressures, \Tc of \FeSeS exhibits a similarly dramatic increase as observed in FeSe \cite{Tomita2012,Tomita2015,Matsuura2017}. Matsuura et al. \cite{Matsuura2017} report that the magnetic dome observed in FeSe above \SI{0.9}{\giga\pascal} \cite{Bendele2010,Bendele2012} is still present but is shifted to higher pressures with increasing S content. Yip et al. \cite{Yip2017} report a local maximum in \Tc at low pressures and a significant weakening of the superconducting diamagnetic shielding that coincides with the verge of the high pressure magnetic dome. Xiang et al. \cite{Xiang2017} conclude from resistivity measurements the onset of an additional small magnetic dome at low pressures at the local maximum in \Tcp. Combining the above results from bulk technique measurements leads to the picture of two well separated magnetic domes in the temperature-pressure phase diagram of \FeSeS which increasingly separate for higher x \cite{Matsuura2017,Xiang2017}. However, local probe muon spin rotation and relaxation ($\mu$SR) measurements, which played a fundamental role in establishing the phase diagram of FeSe under pressure \cite{Bendele2010,Bendele2012}, are missing so far.

In this Letter we present a study of the magnetic and superconducting properties of \FeSeS under hydrostatic pressures up to \SI{2.3}{\giga\pascal} using a combination of muon spin rotation and relaxation ($\mu$SR) \cite{Yaouanc2011}, AC-susceptibility (ACS) and DC-magnetization measurements. It is found that \FeSeSd with an average x~=~0.11 exhibits long range magnetic order above \SI{0.6}{\giga\pascal} with a transition temperature \TN larger than the superconducting transition temperature \Tcp. This magnetic dome is much more extended than the one previously reported for small pressures \cite{Xiang2017} and probably spans all the way to the high pressure magnetic phase \cite{Matsuura2017}. The magnetic order initially competes with the microscopically coexisting superconducting phase leading to a local maximum and minimum of $T_\mathrm{c}(p)$. DC-magnetization measurements on a series of \FeSeS single crystals with x up to 0.14 show that the maximum of \Tc is shifted to lower pressures for increasing sulfur content. This suggests the possibility of magnetism and superconductivity coexisting at ambient pressure for a sample with x~$\geq0.2$.

Five batches of superconducting iron rich \FeSeSd with sulfur content ranging between x~=~0.07 and 0.14 (x~=~0.11 mass weighted average) were grown using the vapor-transport technique \cite{Böhmer2016,FeSeSsupplement}. The batches were powderized and mixed in order to get the minimal sample mass of \SI{1}{\gram} required for \muSR measurements under pressure. DC-magnetization measurements were performed on four different batches of high quality \FeSeS single crystals with well defined $\text{x}=0$, 0.05, 0.09, and 0.14 synthesized and characterized following Ref. \cite{Chareev2018}.
$\mu$SR measurements were performed at the Swiss Muon Source S$\mu$S \cite{Amato2017,Khasanov2016d}.
The data were analyzed with the free software package \textsc{musrfit} \cite{Suter2012}. DC-magnetization measurements were performed using a commercial superconducting quantum interference device (SQUID) magnetometer. Hydrostatic pressure for the \muSR and ACS measurements was applied using double-wall piston cells developed and regularly used at S$\mu$S \cite{Khasanov2016d,Shermadini2017, Bendele2010, Bendele2012, Holenstein2016}. A commercial CuBe piston cell was used for DC-magnetization measurements. Pressures were determined by either In or Sn manometers and Daphne 7373 oil was used as pressure transmitting medium.

\begin{figure}[t]
\centering{
\includegraphics[width=1.0\columnwidth]{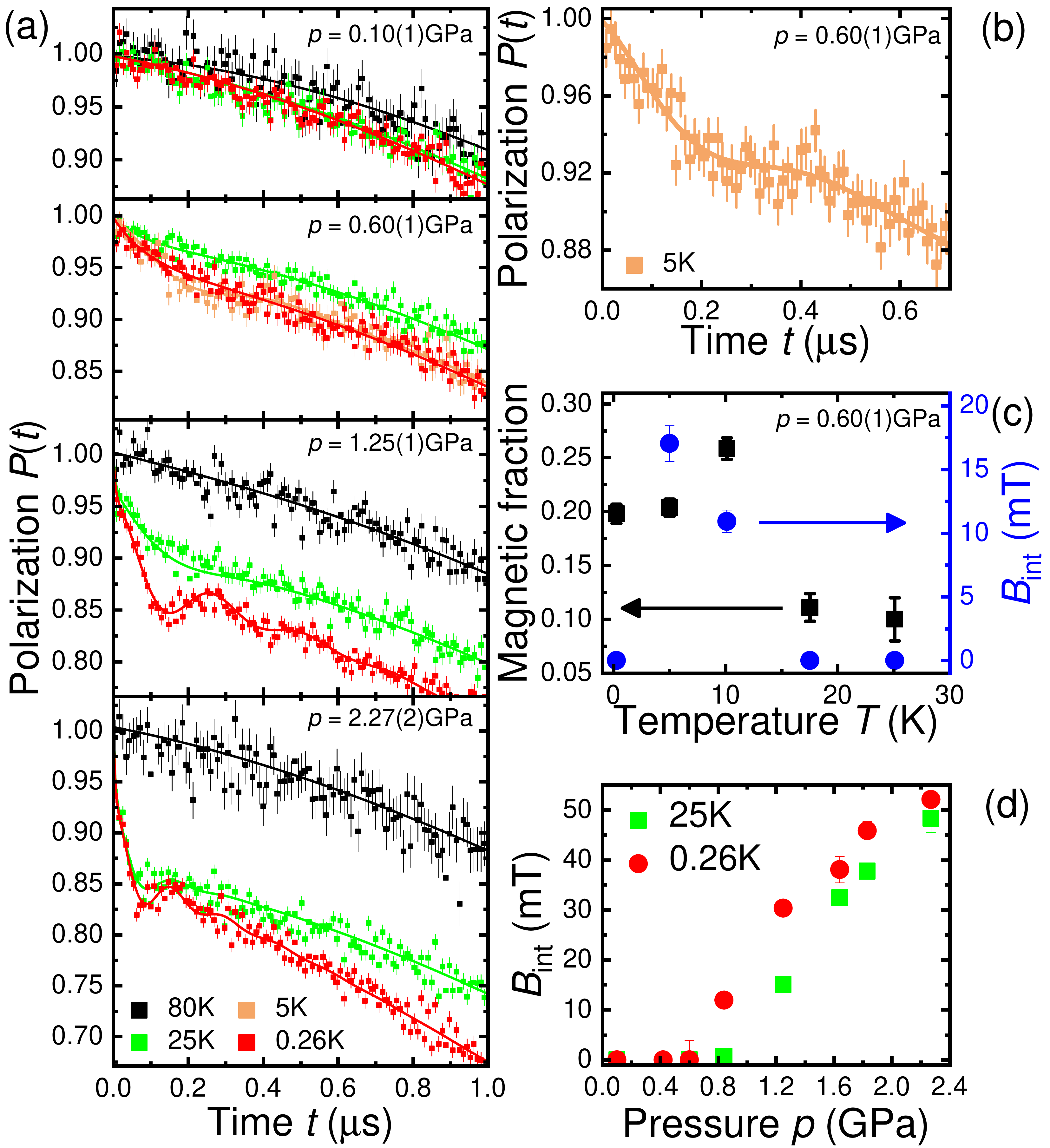}
 \caption{(a), (b) Representative zero-field \muSR spectra of \FeSeSd with a mass weighted average $\text{x}=0.11$ for various pressures and temperatures. The oscillations appearing for pressures above \SI{0.6}{\giga\pascal} indicate the emergence of long range magnetic order. The solid lines are fits using the model detailed in the Supplemental Material \cite{FeSeSsupplement}. (c) Magnetic fraction $f_\mathrm{m}$ and internal field $B_\mathrm{int}$ at the muon stopping site as a function of temperature at \SI{0.6}{\giga\pascal}. The reduction of both quantities towards lower temperatures is due to the coexistence and competition on an atomic scale between the superconducting and the magnetic order. (d) Evolution with pressure of the internal field at 0.26 and \SI{25}{\kelvin}. $B_\mathrm{int}$ is proportional to the local ordered magnetic moment.
 }\label{Fig:Overview}
}
\end{figure}
\begin{figure}[t]
\centering{
\includegraphics[width=0.9\columnwidth]{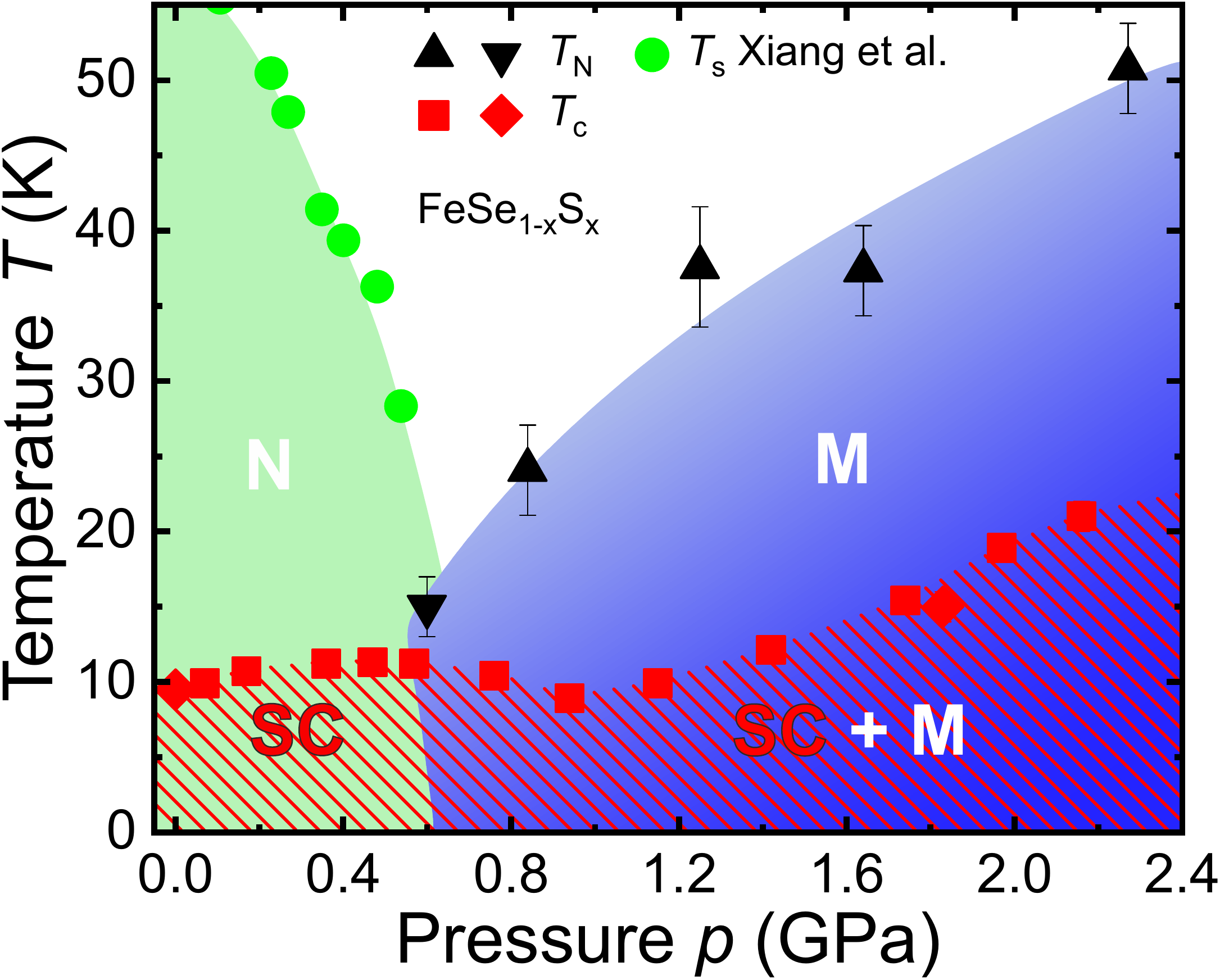}
 \caption{Temperature-pressure phase diagram of \FeSeSd with a mass weighted average $\text{x}=0.11$. The structural transition temperatures $T_\mathrm{s}$ indicating the nematic order (N) are taken from Ref. \cite{Xiang2017} for a \FeSeS sample with $\text{x}=0.096$. The magnetic onset temperature \TN was determined by \SI{5}{\milli\tesla} transverse-field (up facing black triangles) and zero-field (down facing black triangle) \muSR. The superconducting onset temperature \Tc was measured by AC-susceptibility (red squares) and \SI{30}{\milli\tesla} transverse-field \muSR (red diamonds). At \SI{0.6}{\giga\pascal}, long range magnetic order (M) sets in. The competition on an atomic scale between the magnetic and the coexisting superconducting phase (SC+M) at the onset of magnetism leads to a local maximum and minimum in $T_\mathrm{c}$.
 }\label{Fig:Diagram}
}
\end{figure}

The magnetic properties of the \FeSeSd sample under hydrostatic pressure were determined with zero-field (ZF) and \SI{5}{\milli\tesla} transverse-field (TF) \muSRp. Representative ZF muon spin polarization spectra $P(t)$ are shown in Fig.~\ref{Fig:Overview}(a) and (b) for different pressures $p$ and temperatures $T$. At $p=\SI{0.6}{\giga\pascal}$, the onset of spontaneous muon spin precession can be observed below $T_\mathrm{N}\approx\SI{15}{\kelvin}$, which is a clear sign for the emergence of long range magnetic order. The precession frequency is related to the internal magnetic field at the muon stopping site by $\omega_\mathrm{osc}=\gamma_{\mu}B_\mathrm{int}$, where $\gamma_{\mu}=2\pi\times \SI{135.5}{\mega\hertz\per\tesla}$ is the muon's gyromagnetic ratio. A measurement at $p=\SI{0.4}{\giga\pascal}$ (not shown) does not show any sign of magnetic order. A quantitative analysis as detailed in the Supplemental Material \cite{FeSeSsupplement} allows extracting the magnetic volume fraction $f_\mathrm{m}$ and the internal magnetic field $B_\mathrm{int}$ at the muon stopping site. $B_\mathrm{int}$ is proportional to the local ordered magnetic moment (on the Fe atom) and is therefore a measure for the magnetic order parameter. Note that this local probe measurement is therefore qualitatively different from a scattering technique where only the product of magnetic moment and volume can be measured. Interestingly, the locally measured ordered magnetic moment and the magnetic volume fraction do not increase monotonically with decreasing temperature at \SI{0.6}{\giga\pascal} [Fig. \ref{Fig:Overview}(c)]. They reach a maximum between 5 and \SI{10}{\kelvin} and then decrease again towards lower temperatures. A reduction of the ordered magnetic moment size below \Tc was already observed by \muSR for FeSe \cite{Bendele2010,Bendele2012} and by \muSR and Moessbauer spectroscopy for other iron based superconductors \cite{Materne2015,Bernhard2012,Marsik2010,Wang2011b,Wiesenmayer2011}. This is a well established hallmark for a coexistence and competition on an atomic scale. In other words, the ordered magnetic moment of \FeSeS is reduced below \Tc when magnetism and superconductivity compete for the same electronic states in momentum space. Further, in some part of the sample the superconducting order obviously completely suppresses the magnetic order. This results in a reduction of the magnetic volume fraction below \Tcp, in analogy to FeSe [Fig. \ref{Fig:Overview}(c)].
The evolution of $B_\mathrm{int}$ with pressure is summarized in Fig. \ref{Fig:Overview}(d) for 0.26 and \SI{25}{\kelvin}. Compared to FeSe, $B_\mathrm{int}$ is approx. \SI{15}{\percent} smaller at the highest pressure although the magnetic order sets in at a lower pressure \cite{Bendele2012}. The magnetic fraction $f_\mathrm{m}$ increases to $\sim\SI{65}{\percent}$ for the highest pressure currently reachable with \muSRp. To determine the magnetic transition temperature \TN as a function of pressure, \SI{5}{\milli\tesla} TF \muSR was employed \cite{FeSeSsupplement}. The result is shown in Fig. \ref{Fig:Diagram}. Our data indicate that a local magnetic probe such as \muSR observes a much more extended magnetic dome compared to the small dome previously reported at low pressures \cite{Xiang2017}. The dome reported here is likely to span the full pressure range up to the high pressure magnetic phase at $p\approx\SI{4}{\giga\pascal}$ \cite{Matsuura2017}.

\begin{figure}[!b]
\centering{
\includegraphics[width=0.9\columnwidth]{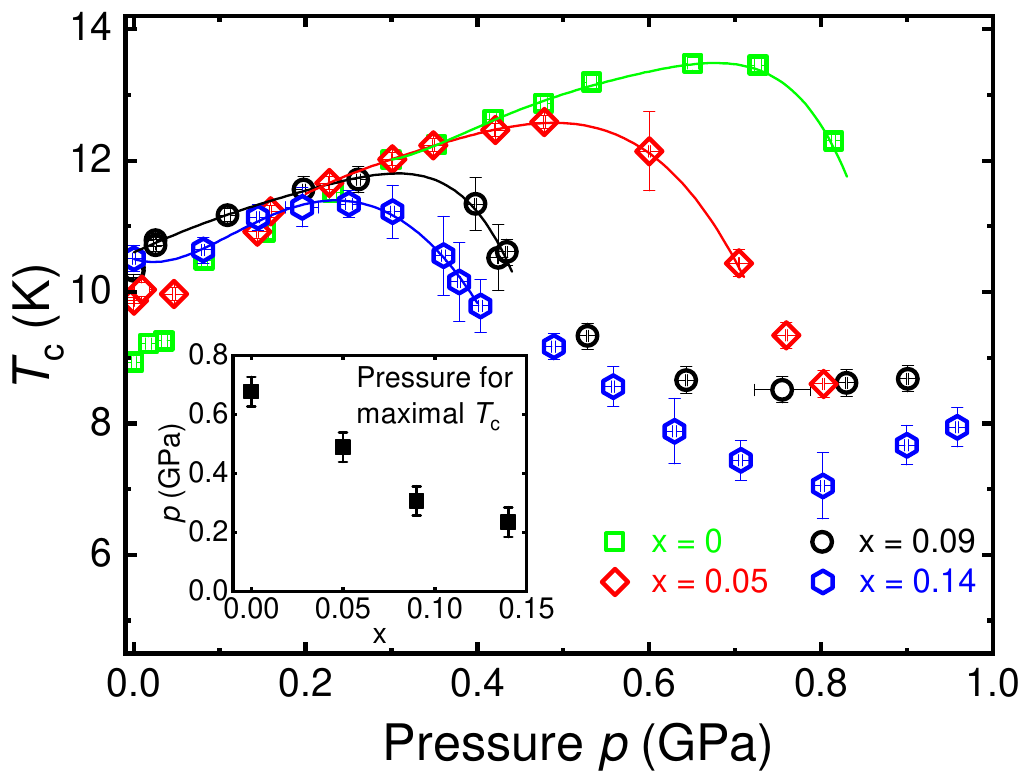}
 \caption{Pressure dependence of the superconducting onset transition temperature \Tc for four different batches of \FeSeS single crystals. The solid lines are guides to the eye. Inset: Change of pressure for the maximal \Tc as a function of sulfur content x.}\label{Fig:SC}
}
\end{figure}

The superconducting properties of \FeSeSd were investigated using TF \muSRp, ACS and DC-magnetization measurements under pressure. At ambient pressure, reported values of the magnetic penetration depth $\lambda$ obtained via the lower critical field $H_\mathrm{c1}$ show a decrease of the superfluid density $n_\mathrm{s}\propto\lambda^{-2}$ for increasing x despite the increase in \Tc \cite{Abdel-Hafiez2015,Mizuguchi2009c}. This is explained with the merging of two superconducting gaps \cite{Abdel-Hafiez2015,Moore2015}. Scanning tunneling microscopy and spectroscopy measurements further emphasize the importance of the multiband character of \FeSeS \cite{Moore2015, Giorgio2016}. Using a two-band model to fit our \muSR relaxation rates \cite{Brandt1988,Serventi2004,Yaouanc2011} for different fields at \SI{0.26}{\kelvin} and ambient pressure \cite{FeSeSsupplement} gives an estimated penetration depth value that is comparable to the ones reported for FeSe \cite{Khasanov2008i,Khasanov2010a,Abdel-Hafiez2015}. Within the accuracy of our measurement we therefore cannot confirm a significant change in superfluid density with S substitution.
The red diamonds in Fig.~\ref{Fig:Diagram} represent the superconducting transition temperature \Tc determined by \SI{30}{\milli\tesla} TF-\muSRp. A more detailed pressure dependence of $T_\mathrm{c}(p)$ was obtained with ACS (red squares in Fig.~\ref{Fig:Diagram}). Also shown in Fig.~\ref{Fig:Diagram} are literature values for the structural transition temperature $T_\mathrm{s}$ of a sample with $\text{x}=0.096$ \cite{Xiang2017}. $T_\mathrm{c}(p)$ slightly decreases above \SI{0.6}{\giga\pascal} where magnetic order sets in, leading to a local maximum and minimum in $T_\mathrm{c}(p)$. This was observed before in FeSe \cite{Bendele2010,Bendele2012} and \FeSeS \cite{Xiang2017,Yip2017}. In the latter case, however, the local maximum in $T_\mathrm{c}(p)$ was attributed to the onset of only a small magnetic dome restricted to the low pressure region and disconnected from the high pressure magnetic phase around $p\approx\SI{4}{\giga\pascal}$ \cite{Xiang2017,Matsuura2017}. Magnetic shielding as observed by ACS \cite{FeSeSsupplement} gets slightly reduced for pressures around the onset of magnetic order at \SI{0.6}{\giga\pascal}. It is however very similar for low ($<\SI{0.6}{\giga\pascal}$) and high ($>\SI{1.5}{\giga\pascal}$) pressures. Combined with the ZF \muSR results [Fig.~\ref{Fig:Overview}(c)] this leads to a picture of initial competition between the magnetic and superconducting order that evolves into a non-competitive microscopic coexistence at higher pressures where \TN and \Tc increase simultaneously. Hence, the local maximum in \Tc indicates the onset of magnetic order. Such a local maximum is also observed by our DC-magnetization measurements under pressure (Fig. \ref{Fig:SC}) on four batches of high quality \FeSeS single crystals with well defined $\text{x}=0$, 0.05, 0.09, and 0.14 \cite{FeSeSsupplement}. With increasing x, the pressure of the local maximum $p_{T_\mathrm{c}^{max}}$(x) decreases (inset Fig. \ref{Fig:SC}). Similar behavior was reported in Refs. \cite{Xiang2017, Yip2017}. Xiang et al. \cite{Xiang2017} have also investigated the x-dependence of the pressure $p_\mathrm{s}(x)$ where the structural transition vanishes. As a function of x, $p_\mathrm{s}(x)$ and $p_{T_\mathrm{c}^{max}}$(x) change at distinctly different rates. Therefore, it is clearly the coexistence and competition of magnetism and superconductivity rather than the suppression of the nematic order which governs the non-monotonic variation of \Tcp(p).

The phase diagram presented in Fig. \ref{Fig:Diagram} looks qualitatively very similar to the one of FeSe \cite{Sun2016,Bendele2010,Bendele2012} in many respects. With increasing pressure, the structural phase transition gets suppressed while \Tc is moderately increased. At the critical pressure $p_\mathrm{c}$, long range magnetic order sets in and \Tc exhibits a local maximum. The magnetic order initially competes with the superconducting order leading to a reduction of \Tcp. After a local minimum in \Tc the competition vanishes as \TN and \Tc increase simultaneously with pressure. However, there are also significant differences between \FeSeS and FeSe. The critical pressure $p_\mathrm{c}$ is $\sim\SI{0.3}{\giga\pascal}$ lower than in FeSe for $x\approx 0.11$. The rates of increase with pressure of the magnetic transition temperature \TN and the ordered magnetic moment are smaller though. In contrast, the rate of increase in \Tc is bigger. This means that the phase diagram of \FeSeS is not just shifted to lower pressures with respect to FeSe but also experiences some significant modifications. This can be phenomenologically explained by the fact that isovalent substitution is not fully equivalent to hydrostatic pressure. Matsuura et al. \cite{Matsuura2017} have shown that the reduction of the lattice constants by little less than \SI{10}{\percent} sulfur substitution is comparable to the one for \SI{0.3}{\giga\pascal} hydrostatic pressure. This matches well with our observed shift in the onset pressure $p_\mathrm{c}$ of magnetism.
The chalcogen height however, which is known to have a significant effect on superconductivity \cite{Okabe2010,Tomita2015,Mizuguchi2010}, gets increased by pressure and decreased by S substitution. Additionally, the different electronegativities of the different chalcogenides lead to a change in the density of states near the Fermi level \cite{Sun2017a}.
In this regard, future measurements with higher sulfur content are highly desirable to check whether the magnetic dome can be shifted all the way down to ambient pressure like indicated for thin films of \FeSeS \cite{Nabeshima2018}. The shift of the local maximum in $T_\mathrm{c}$ (inset of Fig. \ref{Fig:SC}) implies the possibility of a magnetic and superconducting sample at ambient pressure for x~$\geq0.2$. Such a sample would allow more in-depth investigations of the here observed magnetism coexisting with superconductivity. This might shine more light on the influence or lack thereof of magnetism and nematicity on superconductivity.

An important question to discuss is why previous studies on \FeSeS did not detect the extended dome of magnetic order above \SI{0.6}{\giga\pascal} reported in this work. A general problem, especially for measurements under pressure, is certainly the small value of the ordered moment (order of magnitude of tenths of $\si{\micro_{B}}$ per iron for FeSe \cite{Bendele2012,Khasanov2017}). In the case of FeSe, initial studies using, for example, electrical resistivity measurements and $^{57}$Fe Moessbauer spectroscopy \cite{Medvedev2009d} or $^{77}$Se-NMR \cite{Masaki2009} failed to detect static magnetic order. In fact, it was a \muSR study that for the first time unambiguously detected the magnetic order \cite{Bendele2010} with Moessbauer spectroscopy and NMR following much later \cite{Kothapalli2016,Wang2016}. Understandably, non-volume-sensitive techniques like electrical resistivity measurements \cite{Xiang2017,Matsuura2017} can easily miss transitions that do not affect the full volume of the sample and that might additionally be rather broad. More challenging to explain are the difficulties in detecting the magnetic transition for local probes like $^{77}$Se-NMR \cite{Kuwayama2019}. Two possible explanations are: 1) A magnetic structure that produces fields which nearly cancel at the Se sites. 2) The different time windows of NMR and \muSRp. In the latter case, the magnetic structure would be static on the time scale of \muSR but still dynamic for NMR. Interestingly, in FeSe, the second-order structural transition associated with nematic order is suppressed with pressure but a first order structural transition emerges at $p\approx\SI{1.5}{\giga\pascal}$ and persists up to \SI{5.8}{\giga\pascal} \cite{Kothapalli2016, Bohmer2018}. In \FeSeS however, a structural phase transition was found at ambient and high pressures (\SI{4.9}{\giga\pascal}, where a magnetic dome was detected with electrical resistivity measurements) but not in the intermediate pressure region \cite{Matsuura2017,Kuwayama2019}. As we observe magnetism in the intermediate pressure region, this might imply different types of magnetic order. One that is related to the first order structural transition and measurable with electrical resistivity and another one that exists without structural transition and is more difficult to detect. The first order structural transition might be present in a small pressure region at low pressures in FeSe$_\mathrm{1-x}$S$_\mathrm{x}$, too. This could explain the small magnetic dome reported in Ref. \cite{Xiang2017}.
Finally, we discuss the role of sample quality. The sample used for the presented \muSR measurements has a sulfur content ranging between $\text{x}=0.07$ and 0.14 and contains some excess iron. This is however unlikely to explain the different findings. Excess iron was also present in early studies of FeSe \cite{Hsu2008,Bendele2010,Bendele2012}, but has not negated the validity of local probe results. The spread in sulfur content can broaden features like the local maximum in \Tc due to a distribution of transition temperatures and their pressure dependencies. But it cannot account for an overall shift of such a feature. Furthermore, the existence and systematic shift of the local maximum in $T_\mathrm{c}(p)$ is confirmed by our measurements on four high-quality single crystals with different x (Fig. \ref{Fig:SC}) as well as by literature data \cite{Xiang2017,Yip2017}.

In conclusion, we have shown that \FeSeSd with an average x~=~0.11 exhibits a dome of long range magnetic order above a pressure $p\approx\SI{0.6}{\giga\pascal}$. This magnetic phase extends over the intermediate pressure region between the previously reported low and high pressure magnetic domes \cite{Xiang2017, Matsuura2017}. Further, the magnetic order initially competes with the microscopically coexisting superconducting phase. This leads to a local maximum in \Tc where the magnetic order sets in and a local minimum in \Tc where the competition is strongest. At higher pressures ($>\SI{1.5}{\giga\pascal}$), no competition is found. For increasing sulfur content, the local maximum in \Tcp, which coincides with the onset of magnetic order, shifts to lower pressures, roughly extrapolating to a putative superconducting and magnetic \FeSeS sample at ambient pressure for x~$\geq0.2$. The availability of such systems would spur detailed investigations of this newly found magnetism as well as of its interplay with nematicity and its relevance for superconductivity.


\begin{acknowledgments}
This work is partially based on experiments performed at the Swiss Muon Source S$\mu$S, Paul Scherrer Institute, Villigen, Switzerland. We gratefully acknowledge the financial support of S.H. and G.S. by the Swiss National Science Foundation (SNF-Grants No. 200021-159736, No. 200021-149486, and No. 200021-175935), of Z.S. by Horizon 2020 (INFRADEV Proposal No. 654000 “World class Science and Innovation with Neutrons in Europe 2020 (SINE2020)”), and of V.G. by the German Research Foundation (GR 4667/1). D.A.C. is thankful for support by the program 211 of the Russian Federation Government (RFG), Agreement No. 02.A03.21.0006 and by the Russian Government Program of Competitive Growth of Kazan Federal University.
\end{acknowledgments}



%

\widetext
\clearpage
\begin{center}
\textbf{\large Supplemental Material - Extended Magnetic Dome Induced by Low Pressures in Superconducting FeSe$_{1\text{-}x}$S$_{x}$}
\end{center}

\setcounter{equation}{0}
\setcounter{figure}{0}
\setcounter{table}{0}
\makeatletter
\renewcommand{\theequation}{S\arabic{equation}}
\renewcommand{\thefigure}{S\arabic{figure}}
\renewcommand{\bibnumfmt}[1]{[S#1]}
\renewcommand{\citenumfont}[1]{S#1}

\section{Analysis of zero-field \muSR data}

For a quantitative analysis the zero-field (ZF) spectra were fitted taking into account the contribution of the signals coming from muons stopping in the pressure cell and in the sample: $P(t)= f_\mathrm{cell}P_\mathrm{cell}(t)+(1-f_\mathrm{cell})P_\mathrm{sample}(t)$. The signal coming from the pressure cell $P_\mathrm{cell}(t)$ was modelled following Refs. \cite{Khasanov2016dS, Shermadini2017S}. By fixing the corresponding relaxation rates to the literature values, $f_\mathrm{cell}=0.65$ was determined. This is an expected value given the relatively small sample volume. The signal originating from the sample $P_\mathrm{sample}(t)$ has a magnetic and a non-magnetic component:

\begin{equation}\label{Eq:ZF-fitA}
  P_\mathrm{sample}(t)=f_{m}P_\mathrm{magn}(t)+(1-f_{m})P_\mathrm{non\text{-}magn}(t)\,,
\end{equation}
where
\begin{equation}\label{Eq:ZF-fitB}
\begin{split}
  P_\mathrm{magn}(t)=& f_\mathrm{osc}[\frac{2}{3}j_0(\gamma_{\mu}B_\mathrm{int}t)e^{-\lambda_\mathrm{osc}t}+\frac{1}{3}] \\
     & + (1-f_\mathrm{osc})[\frac{2}{3}e^{-\lambda_{m}t}+\frac{1}{3}]\,,
\end{split}
\end{equation}
\begin{equation}\label{Eq:ZF-fitC}
  P_\mathrm{non\text{-}magn}(t)=\frac{2}{3}e^{-\lambda_\mathrm{nm}t}+\frac{1}{3}\,.
\end{equation}

The 2/3 relaxing and 1/3 non relaxing components in the magnetic [$P_\mathrm{magn}(t)$] and non-magnetic [$P_\mathrm{non\text{-}magn}(t)$] contributions to the signal are a consequence of the powder average of the internal fields with respect to the initial muon spin direction in a polycrystalline sample. The magnetic signal itself has to be split again into two parts. One part shows coherent oscillations ($f_\mathrm{osc}$) and is best described by a zeroth order Bessel function $j_0$ as in the case of FeSe \cite{Bendele2010S}. The second part (1-$f_\mathrm{osc}$) is very fast relaxing ($\lambda_{m}\geq\SI{200}{\per\micro\second}$) and likely originates from broad field distributions in regions with static but not yet long range ordered moments. The product $\gamma_{\mu}B_\mathrm{int}$ describes the angular precession frequency of the muon spins, where $B_\mathrm{int}$ is the local field at the muon stopping site and $\gamma_{\mu}=2\pi\times\SI{135.5}{\mega\hertz\per\tesla}$ is the muon's gyromagnetic ratio. $\lambda_\mathrm{osc}$ is about two and $\lambda_\mathrm{nm}$ about three orders of magnitude smaller than $\lambda_{m}$.

\section{Magnetic transition temperature measurement}

\begin{figure}
\includegraphics[width=0.85\columnwidth]{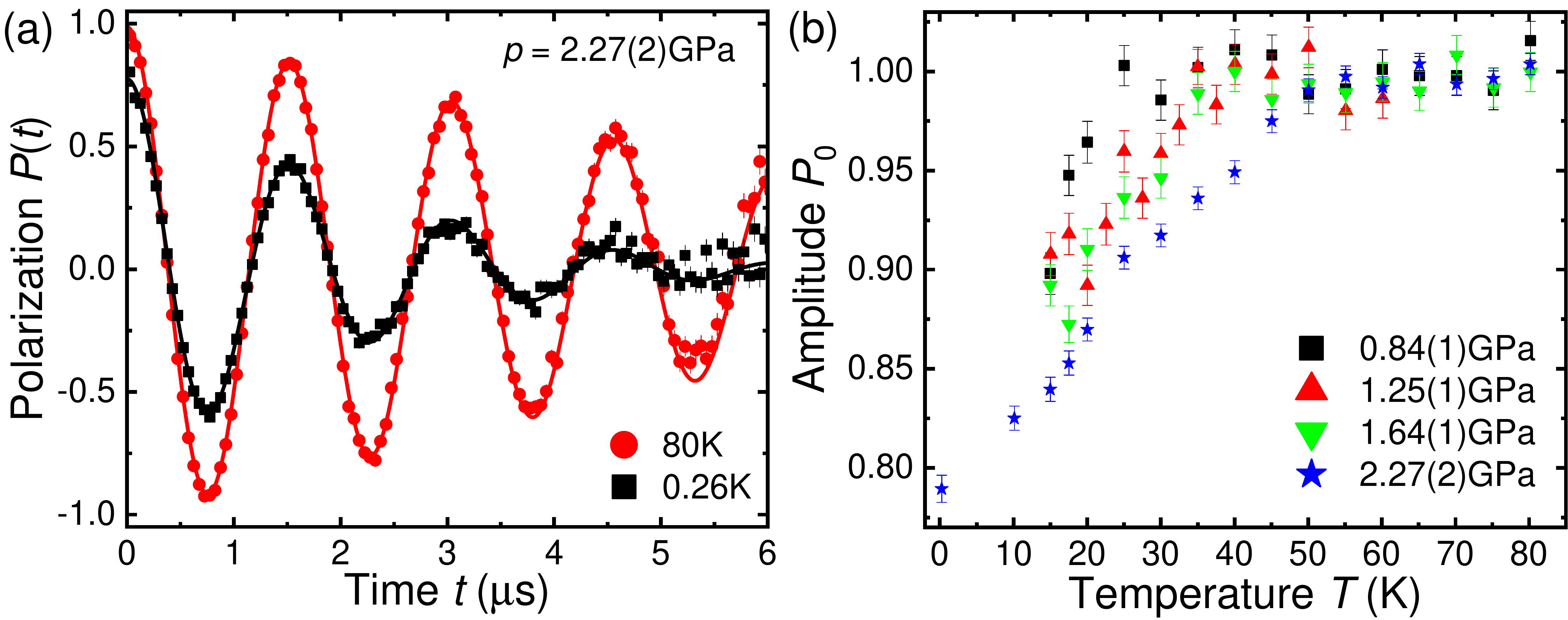}
\caption{(a) Representative \SI{5}{\milli\tesla} TF \muSR spectra above (\SI{80}{\kelvin}) and below (\SI{0.26}{\kelvin}) the magnetic transition. Solid lines are a fit using equation (\ref{Eq:TF}). (b) Amplitude $P_0$ defined in equation (\ref{Eq:TF}) as a function of temperature for different pressures.}\label{TF}
\end{figure}

Transverse-field (TF) muon spin rotation and relaxation (\muSRp) was used to determine the magnetic transition temperature \TNp. An external field of $B=\SI{5}{\milli\tesla}$ was applied perpendicular to the initial direction of the muon spin polarization $P$. This leads to an oscillation of the polarization with an angular frequency $\gamma_\mu B$, where $\gamma_{\mu}=2\pi\times\SI{135.5}{\mega\hertz\per\tesla}$ is the muon's gyromagnetic ratio [see Fig. \ref{TF}(a)]. Magnetic order leads to a very fast depolarization in the first tenths of \si{\micro\second}. The spectra were fitted with a simple phenomenological function, explicitly not catching the fast depolarization:
\begin{equation}\label{Eq:TF}
  P(t)=P_0\cos(\gamma_\mu Bt+\phi)e^{-\lambda t}e^{-\frac{1}{2}(\sigma t)^2}\,.
\end{equation}
Since the muon spin polarization is always 1 when the muons enter the spectrometer, the difference $f=1-P_0$ corresponds to the signal rapidly depolarized by the magnetic fraction. Figure \ref{TF}(b) shows $P_0$ as a function of temperature for different pressures. In the present case, most muons stop in the pressure cell and only \SI{35}{\percent} of the signal originates from the sample. Therefore, the fraction of the sample that becomes magnetically ordered is determined by $f_\mathrm{magn}=(1-P_0)/0.35$. At \SI{2.27}{\giga\pascal} and \SI{0.26}{\kelvin} $f_\mathrm{magn,TF}\approx0.6$ which is in reasonable agreement with $f_\mathrm{magn,ZF}\approx0.65$ found by zero-field (ZF) \muSRp. The transition temperature \TN was determined as the intersection of two linear approximations of the data in Fig. \ref{TF}(b) above and below the transition.

\newpage
\section{Penetration depth measurement}

\begin{figure}
\includegraphics[width=0.5\columnwidth]{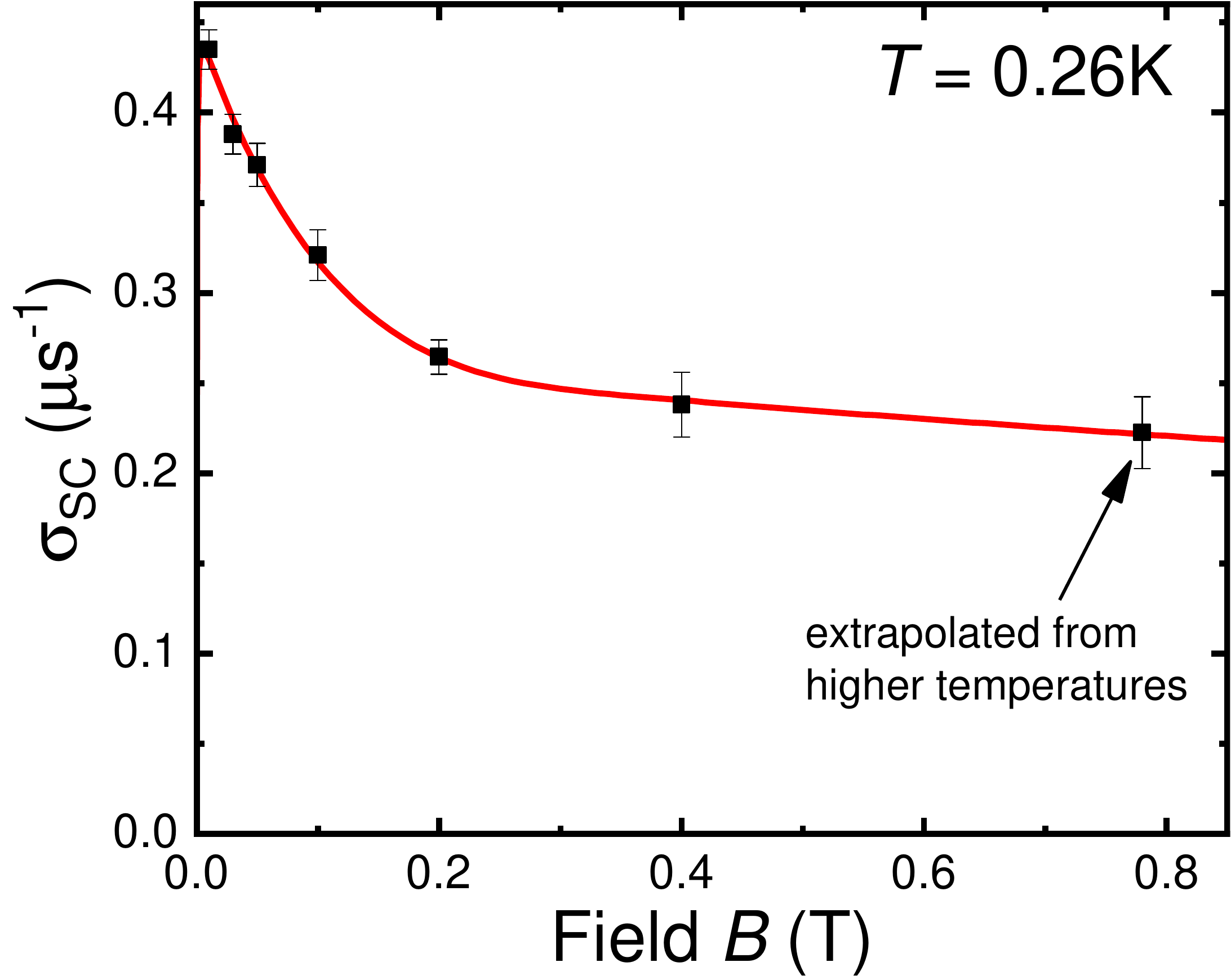}
\caption{Additional Gaussian relaxation rate $\sigma_\mathrm{SC}$ for a batch of \SI{7}{\percent} S substituted FeSe at ambient pressure, \SI{0.26}{\kelvin} and different magnetic fields. The red line is a fit using the two band model introduced in Ref. \cite{Serventi2004S}.}\label{Bscan}
\end{figure}

TF-\muSR is a powerful method to determine the magnetic penetration depth of superconductors. In a field cooled type-II superconductor, the inhomogeneous field distribution of the flux line lattice (FLL) sensed by the muon ensemble leads to an additional Gaussian relaxation $\sigma_\mathrm{SC}$ of the \muSR spectra. The absolute value of the magnetic penetration depth $\lambda$ can be calculated from $\sigma_\mathrm{SC}$ and its field dependence \cite{Brandt1988S,Serventi2004S,Yaouanc2011S}. Fig. \ref{Bscan} shows $\sigma_\mathrm{SC}$ for a batch of \SI{7}{\percent} S substituted FeSe (batch A in Tab. \ref{table}) at ambient pressure, \SI{0.26}{\kelvin}, and different magnetic fields. $\sigma_\mathrm{SC}$ for \SI{0.78}{\tesla} is a value extrapolated from measurements at higher temperatures. The red line is a fit using the two band model introduced in Ref. \cite{Serventi2004S}. This gives a rough estimate for the effective penetration depth $\lambda_\mathrm{eff}\approx\SI{480}{\nano\meter}$ which is a mixture of contributions from the in-plane and out-of-plane penetration depth. The obtained value is very close to the $\lambda_\mathrm{eff}^\mathrm{FeSe}\approx\SI{490}{\nano\meter}$ found for FeSe \cite{Khasanov2010aS}.

\section{AC-susceptibility measurements}

\begin{figure}
\includegraphics[width=0.5\columnwidth]{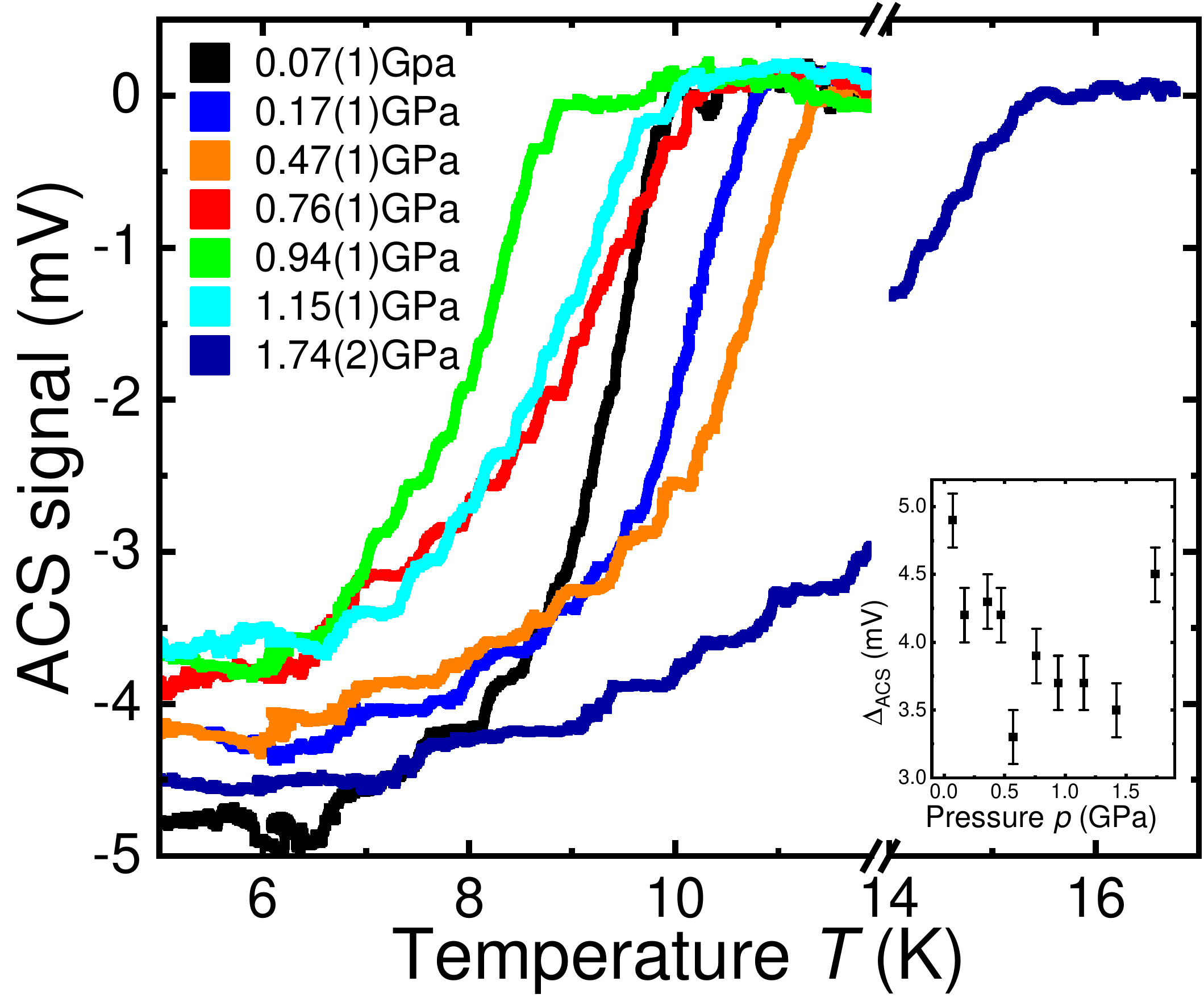}
\caption{Representative AC-susceptibility measurements for different pressures. The superconducting transition temperature changes non-monotonically with pressure. Inset: Change in the ACS signal $\Delta_\mathrm{ACS}$ between the transition and \SI{5}{\kelvin}.}\label{ACS}
\end{figure}

AC susceptibility (ACS) measurements under pressure were performed using a pressure cell designed for \muSR measurements \cite{Khasanov2016dS}. The excitation and pick-up coils were wound around the outside of the cell. This limits the usability of the setup for the investigation of magnetic properties but was employed successfully for the study of superconductors \cite{Bendele2010S,Bendele2012S,Holenstein2016S}. Representative measurements for different pressures are shown in Fig. \ref{ACS}. The superconducting transition temperature \Tc was determined as the intersection of two linear approximations of the data above and below the transition. Apparently, \Tc changes non-monotonically as a function of pressure. Further, the change in the ACS signal $\Delta_\mathrm{ACS}$ between the transition and \SI{5}{\kelvin} is comparable for \SI{0.07}{\giga\pascal} (no magnetic order) and \SI{1.74}{\giga\pascal} (coexistence with magnetic order). The increase in magnetic volume fraction is therefore not at the expense of superconducting volume fraction. This implies a microscopic coexistence of the two orders.

\section{DC-magnetization measurements on high quality single crystals}

\begin{figure}
\includegraphics[width=0.5\columnwidth]{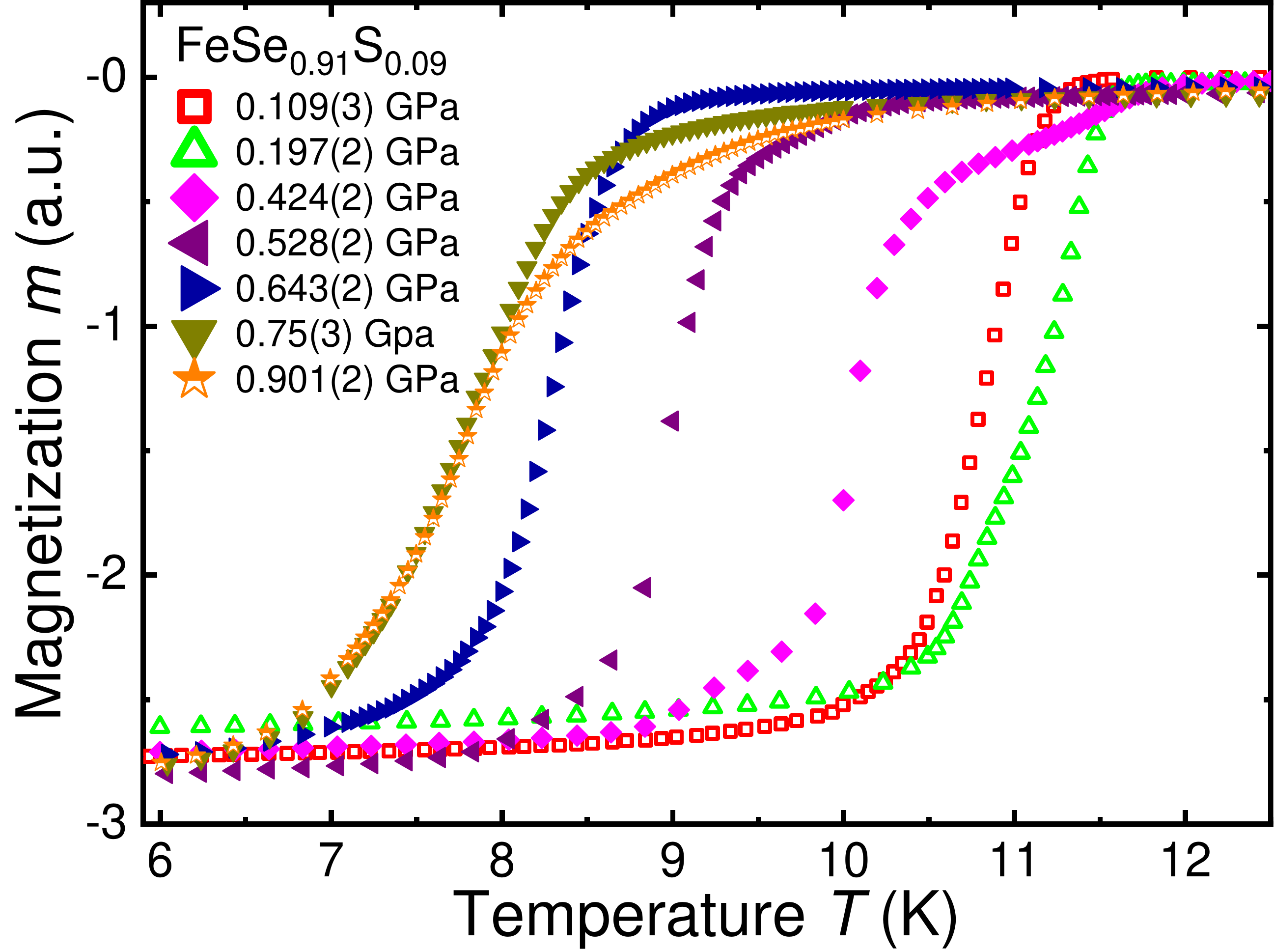}
\caption{Representative DC-magnetization measurements of FeSe$_\mathrm{0.91}$S$_\mathrm{0.09}$ single crystals at different pressures. Like in the polycrystalline sample, the superconducting transition temperature changes non-monotonically with pressure.}\label{DC}
\end{figure}

DC-magnetization measurements under pressure were performed on four batches of high quality \FeSeS single crystals with well defined x = 0, 0.05, 0.09, and 0.14 using a commercial superconducting quantum interference device (SQUID) magnetometer. Hydrostatic pressure was applied by a commercial CuBe piston cell with Daphne 7373 oil as a pressure transmitting medium. Sn was used as a manometer. Figure \ref{DC} shows representative measurements of the magnetization as a function of temperature at different pressures for FeSe$_\mathrm{0.91}$S$_\mathrm{0.09}$ single crystals. Like in the polycrystalline sample used for \muSR measuremets (Fig.~2 of the main text), \Tc changes non-monotonically as a function of pressure (see also Fig.~3 of the main text).

\section{Additional information about the sample used for \muSR and AC-susceptibility measurements}

Samples of \FeSeSd were prepared via the vapor-transport growth technique \cite{Böhmer2016S}. The powdered elements Fe (\SI{456.6}{\milli\gram}; Chempur, \SI{99.9}{\percent}), Se (\SI{513.6}{\milli\gram}; Chempur, \SI{99.999}{\percent}) and S (\SI{29.8}{\milli\gram}; Aldrich, \SI{99.99}{\percent}) with nominal stoichiometry of Fe$_{1.1}$Se$_{0.875}$S$_{0.125}$ were filled together with a mixture of KCl (\SI{2.25}{\gram}, \SI{0.03}{\mole}; Gr\"ussing, \SI{99.5}{\percent}, dried) and AlCl$_3$ (\SI{7.75}{\gram}, \SI{0.06}{\mole}; Alfa Aesar, \SI{99.985}{\percent}) in a glass ampoule of \SI{4}{\centi\meter} length and \SI{5}{\centi\meter} diameter. The ampoule was heated to \SI{390}{\degreeCelsius} at the bottom and 260-\SI{280}{\degreeCelsius} at the top for 5-10 days. After washing with water and ethanol the products were dried under vacuum at room temperature. Lattice parameters were determined by a Rietveld fit of powder X-ray diffraction (PXRD) data. The composition was determined by energy dispersive X-ray (EDX) analysis. All results are summarized in Tab. \ref{table}. The five batches were powderized and mixed in order to get the minimal sample mass of \SI{1}{\gram} required for \muSR measurements under pressure.

\begin{table}[!h]
  \centering
  \renewcommand{\arraystretch}{1.5}
  \renewcommand{\tabcolsep}{0.5cm}
  \begin{tabular}{clll}
  \hline
  Batch & Lattice parameters (\si{\angstrom}) & Composition & Mass (\si{\milli\gram})\\
  \hline
  A & a=3.76127(8), c=5.4919(2) & Fe$_{1.05(3)}$Se$_{0.93(3)}$S$_{0.07(3)}$ & 200 \\
  B & a=3.76678(4), c=5.4991(1) & Fe$_{1.15(7)}$Se$_{0.89(2)}$S$_{0.11(2)}$ & 180 \\
  C & a=3.76659(4), c=5.5015(2) & Fe$_{1.18(7)}$Se$_{0.91(5)}$S$_{0.09(5)}$ & 140 \\
  D & a=3.76650(4), c=5.5005(1) & Fe$_{1.19(6)}$Se$_{0.90(4)}$S$_{0.10(4)}$ & 240 \\
  E & a=3.75882(4), c=5.4764(1) & Fe$_{1.16(6)}$Se$_{0.86(4)}$S$_{0.14(4)}$ & 260 \\
  \hline
  \end{tabular}
  \caption{Overview of samples used for \muSR and AC-susceptibility measurements.}\label{table}
\end{table}




%


\end{document}